\documentclass[floatfix,showpacs,amsmath,amsfonts,amssymb,aps,twocolumn, prb,groupedaddress]{revtex4-1}

\usepackage[table,dvipsnames]{xcolor}
\usepackage{graphicx}
\usepackage{dcolumn}
\usepackage{amsthm}
\usepackage{bm}
\usepackage{siunitx}
\usepackage{nicefrac}
\usepackage{todonotes}
\usepackage{mathtools}
\DeclarePairedDelimiter{\ceil}{\lceil}{\rceil}

\usepackage{float}
\usepackage{braket}
\usepackage{makecell} 
\usepackage{multirow}
\usepackage{wasysym}
\usepackage{titletoc}
\usepackage{gensymb}
\usepackage{multirow}
\usepackage{mwe}
\usepackage{diagbox}

\definecolor{darkred}{rgb}{0.6,0.,0.}
\definecolor{darkgreen}{rgb}{0.,0.5,0.}
\definecolor{darkblue}{rgb}{0.,0.,0.6}
\usepackage[ breaklinks, colorlinks=true, linkcolor=darkred, citecolor=darkgreen, urlcolor=darkblue]{hyperref}
\usepackage{autonum}

\newcommand{\includegraphicsgood}{\includegraphics}

\usepackage[inactive,tightpage]{preview}
\PreviewMacro[*[[!]{\input{}}
\PreviewMacro[*[[!]{\includegraphicsgood}
\newcommand{\normord}[1]{:\mathrel{#1}:}

\begin{document}

\title{Abelian topological order of $\nu=2/5$ and $3/7$ fractional quantum Hall states in lattice models}
\author{Bartholomew Andrews}
\author{Madhav Mohan}
\author{Titus Neupert}
\affiliation{Department of Physics, University of Zurich, Winterthurerstrasse 190, 8057 Zurich, Switzerland}
\date{\today}

\begin{abstract}

Determining the statistics of elementary excitations supported by fractional quantum Hall states is crucial to understanding their properties and potential applications. In this paper, we use the topological entanglement entropy as an indicator of Abelian statistics to investigate the single-component $\nu=2/5$ and $3/7$ states for the Hofstadter model in the band mixing regime. We perform many-body simulations using the infinite cylinder density matrix renormalization group and present an efficient algorithm to construct the area law of entanglement, which accounts for both numerical and statistical errors. Using this algorithm, we show that the $\nu=2/5$ and $3/7$ states exhibit Abelian topological order in the case of two-body nearest-neighbor interactions. Moreover, we discuss the sensitivity of the proposed method and fractional quantum Hall states with respect to interaction range and strength.

\end{abstract}

\maketitle         

\section{Introduction}

Throughout the extensive history of the fractional quantum Hall (FQH) effect, a lot has been learned about the elementary excitations above the ground state at FQH plateaus. In the continuum, numerous trial wavefunctions have been checked using numerical calculations~\cite{Jain90, Morf02} and the elementary excitations of them are well defined~\cite{Jain07, Stern08, Quinn09}. Moreover, Chern-Simons theories can be constructed for many common states and their elementary excitations can also be understood within this framework~\cite{Hansson17, Murthy03}. Building on this, modern research efforts have focused on generalized FQH states in lattice models~\cite{Neupert11, Tang11, Regnault11}, which offer rich physics not present in the continuum~\cite{Liu12, Liu13, Liu13_2} and are currently at the forefront of experimental research~\cite{Spanton18, Nielsen13}. Among the defining features under investigation are the fractional excitations of the ground state, known as anyons, which can obey either Abelian or non-Abelian statistics~\cite{Leinaas77, Wilczek82, Froehlich88, Moore91, Wen91}. In the Abelian case, exchange of quasiparticles in a given ground state yields a fractional phase shift of the wavefunction represented by a one-dimensional braid group, whereas in the non-Abelian case, the ground state is highly degenerate and an exchange of quasiparticles additionally shifts between ground states, which is represented by a higher-dimensional braid group. Such non-Abelian states are particularly intriguing due to their potential application to topological quantum computing~\cite{Nayak08}, as well as other exotic properties~\cite{Sarma15}.  

Continuum FQH states in the Jain hierarchy ($\nu=1/3,2/5,3/7,\dots$) have been shown numerically~\cite{Jain90, Morf02, Toke09}, and for the Laughlin state also experimentally~\cite{Nakamura20, Bartolomei20}, to possess Abelian topological order for the Coulomb interaction. The analysis of corresponding lattice FQH states, on the other hand, is complicated by several factors, including the limited number of viable experimental systems~\cite{Nielsen13}, and the difficulty of engineering long-range interactions~\cite{Nandy19}. Coupled with this, it has been shown that the lattice can host fundamentally different phases of matter~\cite{Bergholtz13}, as well as states with non-Abelian statistics at equivalent filling factors~\cite{Liu13_2, Sterdyniak13}. In particular, there are Abelian FQH states in lattice models stabilized by two-body interactions that have been shown to possess non-Abelian statistics when interactions are sufficiently long-range~\cite{Liu13_2}, which provides motivation for further study. Given the experimental challenges and theoretical interest, it is important to develop an efficient method to analyze such states numerically and probe their quantum statistics.

In this paper, we perform large-scale numerical calculations to investigate the Abelian nature of the single-component $\nu=2/5$ and $3/7$ FQH states in the Hofstadter model with a large interaction strength, chosen such that inter-band transitions are likely to occur. We compute the entanglement entropy at various system sizes to construct detailed plots of the area law of entanglement and we extrapolate to the thermodynamic limit to read off the topological entanglement entropy~\cite{Kitaev06, Levin06} -- an indicator of quasiparticle statistics. We build on previous studies in the field by presenting an efficient algorithm, which addresses both numerical and statistical errors. Using this algorithm, we are able to compute the topological entanglement entropy to a high precision and demonstrate that these states are Abelian in the case of nearest-neighbor interactions. Furthermore, we examine the sensitivity of the procedure and FQH configurations with respect to interaction range and strength. Notably, we find an increase in topological entanglement entropy with interaction range, which leads us to discuss the scope of the algorithm and the fate of these states in the long-range interaction regime. 

This paper is structured as follows. In Sec.~\ref{sec:model} we introduce and justify the kinetic and interaction terms of the Hamiltonian. In Sec.~\ref{sec:method} we outline the numerical method and discuss efforts to minimize the errors in our extrapolation, which is crucial for this investigation. In Sec.~\ref{sec:results} we present the results for the $\nu=1/3$, $2/5$, and $3/7$ states, and finally in Sec.~\ref{sec:discussion}, we discuss the results in the context of current research and highlight promising directions for further study.

\section{Model}
\label{sec:model}

In this section, we introduce the single-particle Hamiltonian in Sec.~\ref{subsec:single} and the corresponding many-body Hamiltonian in Sec.~\ref{subsec:many}.  

\subsection{Single-particle Hamiltonian}
\label{subsec:single}

We consider spinless fermions hopping on a square lattice in the $xy$-plane subject to a perpendicular magnetic field. The single-particle Hamiltonian describing the kinetic energy is given by the Hofstadter model~\cite{Hofstadter76}
\begin{equation}
\label{eq:single_ham}
H_0 = \sum_{\braket{ij}_1} \left[ t e^{\mathrm{i}\theta_{ij}} c_i^\dagger c_j + \text{H.c.} \right],
\end{equation}
where $t$ is the hopping amplitude, $e^{\mathrm{i}\theta_{ij}}$ is the Peierls phase factor, $c^\dagger(c)$ are the creation(annihilation) operators for spinless fermions, and $\braket{ij}_{\kappa'}$ denotes pairs of $\kappa'$th nearest-neighbor sites on the square lattice.

Due to the presence of the perpendicular magnetic field, $\mathbf{B}=B\hat{\mathbf{e}}_z$, the fermions acquire an Aharanov-Bohm phase as they hop around a plaquette~\cite{Aharonov59}. The precise value of the phase along a particular path is dependent on the gauge, since $\theta_{ij}=\int_i^j \mathbf{A}\cdot \mathrm{d}\mathbf{l}$, where $\mathbf{A}$ is the vector potential and $\mathrm{d}\mathbf{l}$ is the infinitesimal line element from sites $i$ to $j$~\cite{Peierls33}. In this paper, we choose to work in $x$-direction Landau gauge such that $\mathbf{A}=Bx\hat{\mathbf{e}}_y$. For this choice of gauge, the fermions only acquire a phase when they hop in the $x$-direction. The periodicity of the phases in the $x$-direction define the magnetic translation algebra~\cite{Zak64} and the unit cell of the system is extended to a magnetic unit cell (MUC) of area $A_\text{MUC}=q\times 1$, where $q$ is an integer and the lattice constant, $a\equiv1$. The dimensions of our system are $L_x\times L_y$ and, unless explicitly stated, are given in units of the corresponding MUC dimensions. Crucially, there are two competing area scales in the model, the irreducible area of a flux quantum and the magnetic unit cell area, which gives rise to a fractal energy spectrum with an infinite selection of Chern bands~\cite{Azbel64}. This frustration is typically quantified using the flux density, defined as $n_\phi=BA_\text{UC}/\phi_0\equiv p/q$, where $\phi_0=h/e$ is the flux quantum and $p, q$ are coprime integers. In the square-lattice Hofstadter model, $q$ directly corresponds to the number of bands in the spectrum and increasing $q$ decreases the band width, such that $n_\phi\to 0$ corresponds to the continuum limit~\cite{Andrews18}. Moreover the filling factor of the lowest Landau level is given as $\nu=n/n_\phi$, where $n$ is the filling factor of the system.

There are several important advantages of using the Hofstadter model for this investigation. First, due to the fractal energy spectrum with any desired Chern band, the system is highly configurable and we can easily access all of the desired topological flat bands. Second, the Hamiltonian is computationally minimal, since it only requires nearest-neighbor hopping on a square lattice with a phase factor, albeit with an enlarged effective unit cell. Third, and crucially for this paper, we can tune the relevant length scale in the problem, the magnetic length, simply by adjusting the flux density. This allows us to access a larger selection of system sizes at a low computational cost. We can also choose the magnetic unit cell dimensions to be in the preferred direction for our algorithm, which in our case is the $x$-direction (as discussed later). This powerful combination of configurability and simplicity make the Hofstadter model an ideal choice for our study.

We note that since the kinetic energy of the system is quenched for FQH states, it is typically the interaction Hamiltonian that dominates the physics. Ideally, the role of the single-particle Hamiltonian is simply to facilitate tuning to the correct system configurations.

\subsection{Many-body Hamiltonian}
\label{subsec:many}

The many-body Hamiltonian comprises the single-particle Hamiltonian (\ref{eq:single_ham}) with the addition of a density-density interaction term, such that
\begin{equation}
\label{eq:many_ham}
H = H_0 + \sum_{\kappa'=1}^{\ceil{\kappa}} V(\kappa', n_\phi) f_{\kappa'-1}(\kappa) \sum_{\braket{ij}_{\kappa'}} \rho_i \rho_j,
\end{equation}
where $\kappa\in\mathbb{R}^+$ is the interaction range, $V(\kappa', n_\phi)=V_0/(\kappa'/l_\text{B})$ is the Coulomb potential accounting for the fact that the lattice constant in units of $l_B$ varies as a function of $n_\phi$, $f_{i}(\kappa)=\min\{\Theta(\kappa-i)(\kappa-i),1\}$ is a scale factor involving the Heaviside step function $\Theta$, and $\rho_i=c^\dagger_i c_i$ is the fermionic density operator. The sum is constructed such that we can tune the interaction range continuously with respect to $\kappa$. In cases where $\kappa$ is non-integer, we scale the $\ceil{\kappa}$th nearest-neighbor term by the fractional part of $\kappa$, where $\ceil{\dots}$ denotes the ceiling operator. The interaction strength constant, $V_0=10$, is chosen such that inter-band transitions are likely for all systems. The density-density form of the interaction term is chosen predominantly due to its simplicity, and hence low computational expense. We note that although non-Abelian FQH states were originally introduced via many-body interaction terms, it has since been shown that equivalent phases may also be stabilized with two-body interactions~\cite{Kuszmierz18, Liu13, Liu13_2}. 

\section{Method}
\label{sec:method}

To solve the many-body problem, we employ the infinite density matrix renormalization group (iDMRG) method on a thin cylinder geometry~\cite{White92, Schollwoeck11, Cincio13}. The iDMRG algorithm works by transcribing an initial wavefunction into a matrix product state (MPS) and the corresponding Hamiltonian into a matrix product operator (MPO). In traditional MPS-based DMRG, there is a matrix assigned to each site of a one-dimensional chain. In this case, since we are modeling two dimensions, the chain snakes to cover the entire system. At each site the matrices representing the wavefunction are truncated up to a MPS bond dimension, $\chi$, such that $\ket{\psi}=\sum_{\alpha=1}^\chi \Lambda_\alpha \ket{\alpha_\text{L}}\otimes\ket{\alpha_\text{R}}$, where we perform a Schmidt decomposition such that the cylinder is spatially cut into left and right halves. $\Lambda_\alpha$ are referred to as the Schmidt coefficients and $\ket{\alpha_{\text{L}/\text{R}}}$ are the left/right Schmidt states. The crucial property for DMRG is that the Schmidt eigenbasis may be directly related to the eigenbasis of the reduced density matrix $\rho^{\text{L}/\text{R}}_\alpha=\text{Tr}_{\text{R}/\text{L}}\rho_\alpha$, and hence provides a way of accessing entanglement properties. In fact, the Schmidt states correspond directly to the eigenstates of the reduced density matrix and the Schmidt values are the square of the eigenvalues. One of the central quantities for the algorithm is the entanglement entropy and spectrum, since this offers insight into topological features. Typically, the von Neumann entanglement entropy is used, defined as $S_\text{vN}=-\sum_{\alpha=1}^\chi \Lambda_\alpha^2 \ln (\Lambda_\alpha^2)$ with $\Lambda_\alpha\equiv \mathrm{e}^{-\epsilon_\alpha/2}$, where $\epsilon_\alpha$ are the entanglement energies. The iDMRG algorithm in this paper is set on an infinite cylinder geometry, where translational invariance is assumed along the cylinder axis ($x$-direction) and periodic boundary conditions are taken along the circumference ($y$-direction). The algorithm sweeps iteratively along the one-dimensional chain, performing two-site updates after relaxing the system in accordance with the Lanczos algorithm and truncating in accordance with the maximum given MPS bond dimension. It continues in this fashion until convergence of the energy, entropy, and other discerning quantities.

For the purposes of this project, the iDMRG algorithm offers both notable advantages, as well as some drawbacks. The main advantage is that, unlike exact diagonalization, no band projection needs to be taken for the interaction Hamiltonian and so inter-band transition effects are automatically taken into account. An added benefit is that the algorithm works in the semi-thermodynamic limit, meaning that a thermodynamic limit ansatz is taken along the cylinder axis. Furthermore, the system sizes attainable along the circumference are highly competitive with alternative methods. The major disadvantage of the algorithm is that it is inherently one-dimensional, which means that even modest interaction ranges on the two-dimension surface correspond to exponentially long-range interactions on the unraveled one-dimensional chain. This is particularly an issue for this project as we are motivated to tune the interaction range. Nevertheless, we overcome this barrier through the use of an optimal sampling algorithm, as explained in later sections.        

The way in which we identify Abelian topological order in this paper is based on the area law of entanglement, $S=\alpha L -\gamma + O(\mathrm{e}^{-L})$, where $\alpha$ is a non-universal constant dependent on the microscopic parameters of the Hamiltonian, $L$ is the system size in a given direction, and $\gamma$ is the topological entanglement entropy~\cite{Kitaev06, Levin06}. The topological entanglement entropy cannot be removed by reducing the system size and depends intrinsically on the type of quasiparticle excitations hosted by the ground state. It is generally written as $\gamma=\ln(\mathcal{D})$, where $\mathcal{D}=\sqrt{\sum_a d_a}$ is the total quantum dimension of the field theory description and $d_a$ is the quantum dimension of a quasiparticle of type $a$. The quantum dimension for Abelian anyons is always one, whereas for non-Abelian anyons $d_a>1$~\cite{Kitaev06}. The conventional Laughlin argument for an Abelian FQH state at filling $\nu=r/s$ is that the ground state degeneracy is $s$, the quasiparticles possess $1/s$ of an electronic charge, and the topological entanglement entropy is $\gamma=\ln(\sqrt{s})$~\cite{Laughlin83, Estienne15}. For non-Abelian order, this value is always larger. We note, however, that it has recently been shown that the topological entanglement entropy may be larger than $\ln(\sqrt{s})$, even for single-component Abelian states, taking the general form $\gamma=\ln(\sqrt{\lambda s})$ with $\lambda\in\mathbb{Z}^+$~\cite{Balram20, Balram20_2}. Therefore, the original statement $\gamma=\ln(\sqrt{s})$ implies that the state is Abelian but the converse is not always true. By precisely extrapolating the value of the topological entanglement entropy from the area law of entanglement, we are able to definitively conclude that a state exhibits Abelian order when $\gamma=\ln(\sqrt{s})$.

Although the premise is simple, the execution is fraught with potential problems. The first and perhaps most apparent problem is the precision to which we are able to extrapolate to the topological entanglement entropy. In many-body numerics, such as iDMRG, we are restricted to relatively small system sizes. Not only is the area law technically non-linear at small system sizes, but more importantly, finite-size effects exist on top of this, leaving the area law data highly spread and unreliable. Second the individual data points are computed at a finite bond dimension, which may be significantly off from the actual value in the $\chi\to\infty$ limit. Even slight errors in the individual data points can have a compound effect on the total error of the topological entanglement entropy, particularly if the points are close together on the $L$-axis. While attempting to alleviate these issues, there is additionally an arbitrariness in how to construct the line of best fit -- which points should be included and which should be left out? Minor changes in acceptance criteria can have a drastic impact on the slope and $y$-intercept of the linear regression. Although a lot has been achieved with such computations in the past few years~\cite{Grushin15, Schoonder19, Andrews20}, we argue that more precise numerics are required to draw reliable conclusions. Specifically, in order to perform a stand-alone computation of the topological entanglement entropy, more care is required to address numerical and statistical errors.

In this paper, we address these issues and present a systematic method that is reliable enough to accurately compute the topological entanglement entropies for the $\nu=1/3$ and $2/5$ states. Moreover, based on this procedure, we also obtain an estimate for the topological entanglement entropy of the $\nu=3/7$ state. First, we plot the area law of entanglement in units of magnetic length, which in the square-lattice Hofstadter model depends on the flux density through $L_y/l_B=\sqrt{2\pi n_\phi}L_y$. Since the processing cost for convergence for iDMRG scales exponentially with $L_y$, exploiting the natural length scale of the Hofstadter model allows us to obtain a larger number of data points at relatively low computational cost~\cite{Schoonder19}. Second, to remove arbitrariness and optimize the flux densities considered, we choose values guided by an algorithm (Sec.~\ref{subsec:nphi}). For each data point, we scale the computation of the entanglement entropy with bond dimension to obtain an extrapolation with error in the $\chi\to\infty$ limit. We accept data points only if the error is smaller than $0.1\%$ (Sec.~\ref{subsec:extrapolation}). Finally, to mitigate finite-size effects, we construct multiple lines of best fit as we incrementally exclude data in ascending $L_y/l_B$, and stop as soon as the linear regression of the remaining points yields $R^2>0.99$. This is a necessary compromise between a precisely straight line and a maximal data set (Sec.~\ref{subsec:line}). The details of the numerical method are discussed in Appendix~\ref{sec:num_method}.     

\section{Results}
\label{sec:results}

In this section we present our results from the many-body numerical calculations. In Sec.~\ref{subsec:NN_int} we demonstrate the Abelian statistics of the FQH states in the case of nearest-neighbor interactions, and in Secs.~\ref{subsec:tune_int_range} and~\ref{subsec:tune_int_strength} we analyze the effect on the topological entanglement entropy as we tune the interaction range and strength, respectively.

\subsection{Nearest-neighbor interactions}
\label{subsec:NN_int}

\begin{figure}
	\includegraphics[width=\linewidth]{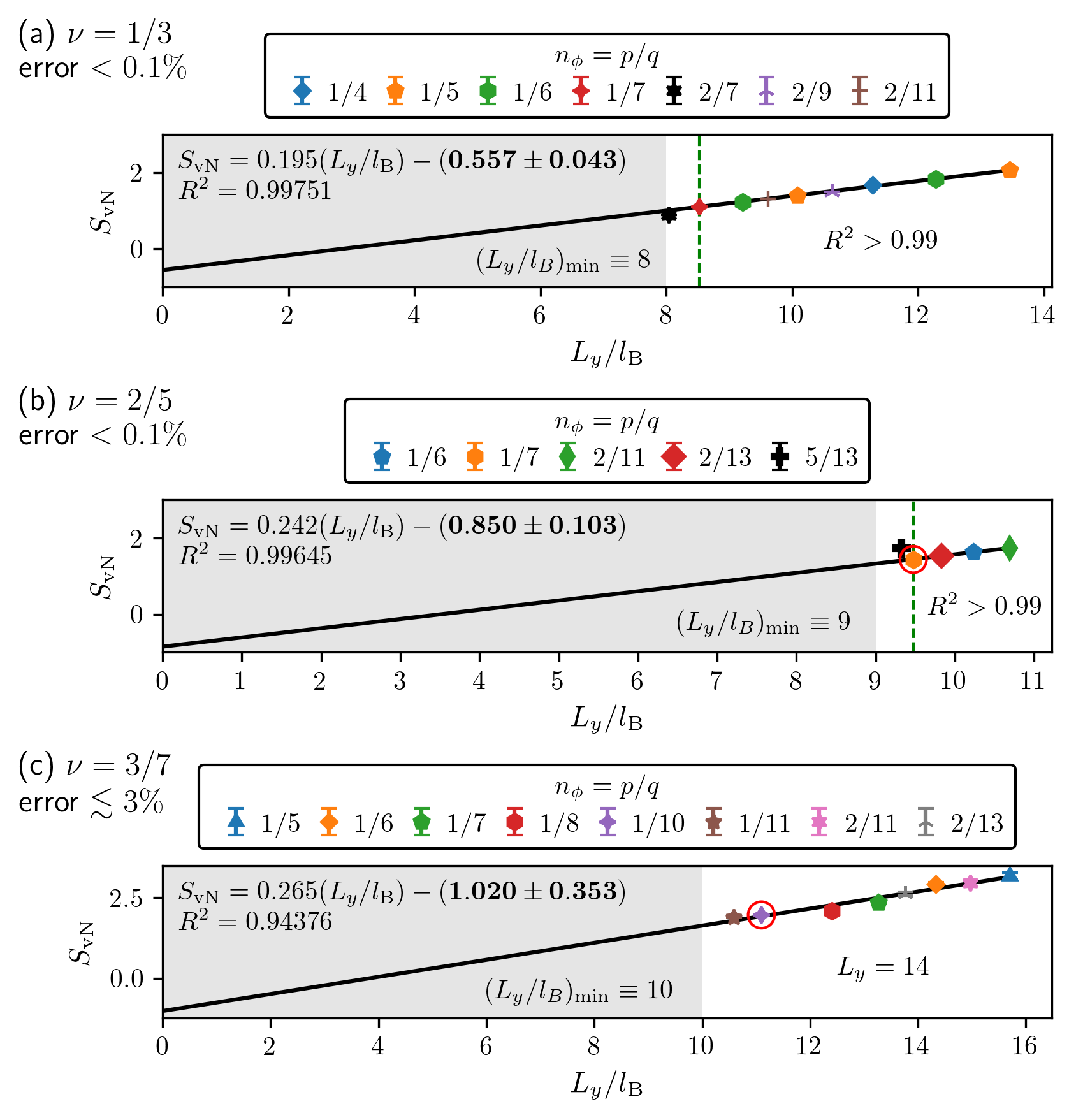}
	\caption{\label{fig:short_range_int}Von Neumann entanglement entropy, $S_\text{vN}$, against cylinder circumference, $L_y$, in units of magnetic length, $l_B=(2\pi n_\phi)^{-1/2}$, for the fermionic hierarchy states at (a) $\nu=1/3$, (b) $\nu=2/5$, and (c) $\nu=3/7$. In each case, we use nearest-neighbor interactions ($\kappa=1$). (a, b) The threshold where the $R^2$ value first exceeds 0.99 is marked with a green dashed line. All of the points above this threshold are used to construct the line of best fit. In (c), we present the complete data set with our largest system size $L_y=14$. In all cases, we obtain points based on the systematic procedure outlined in Appendix~\ref{sec:num_method} with $\chi_\text{max}=3000$. The complete data sets are shown in Fig.~\ref{fig:complete_data} and the points circled in red are studied in Fig.~\ref{fig:case_studies}.}
\end{figure}

To begin, we consider the FQH states stabilized by nearest-neighbor interactions, as defined in the interaction Hamiltonian (\ref{eq:many_ham}) with $\kappa=1$.

To benchmark our results, we start with the Laughlin filling $\nu=1/3$. Although this state has been previously investigated using an area law constructed from a many-body lattice simulation~\cite{Grushin15, Schoonder19, Andrews20}, we emphasize that the cited investigations are not systematic enough to be transferable for higher-order states in the hierarchy. We therefore present the computation of the $\nu=1/3$ area law plot using our systematic procedure in Fig.~\ref{fig:short_range_int}.(a). We have algorithmically chosen our data points to avoid selection bias, we have scaled each data point with $\chi$ to eliminate convergence error, and we have systematically excluded small-$L_y/l_B$ data to alleviate finite-size effects. Most importantly, all points are converged in the entanglement entropy to $\delta S<0.1\%$, have a spacing of $\Delta (L_y/l_B)>0.1$, and the linear regression satisfies $R^2>0.99$ (see Appendix~\ref{sec:num_method}). Further tightening the constraints of the algorithm yields eight such points, shown in the figure. We note that for the $\nu=1/3$ state we obtained significantly more data points that satisfy all of the criteria, which is why we further restricted the algorithm to yield a smaller representative sample (as detailed in Appendix~\ref{sec:complete_data}). The topological entanglement entropy obtained from these data is $\gamma = 0.557\pm0.043$, which agrees closely with the Abelian theory value of $\gamma=\ln(\sqrt{3})\approx 0.549$\footnote{Moreover, this is far from the value taken by the closest non-Abelian competitor, the particle-hole conjugate of the four-cluster Read-Rezayi state, which has a topological entanglement entropy of $\gamma\approx1.792$~\cite{Peterson15}.}.

We progress from the Laughlin state to the next filling fraction in the hierarchy: $\nu=2/5$. We hold the $\nu=2/5$ data to the same stringent quality standards that we enforced for the $\nu=1/3$ state. The data obtained are shown in Fig.~\ref{fig:short_range_int}.(b), which serves as our first original result. Since for the $\nu=2/5$ state it is more difficult to satisfy configuration constraints and convergence at accessible $\chi$, this filling presents a significant computational challenge compared to the Laughlin state\footnote{The difficulty in computing higher filling fractions in the FQH hierarchy is not only due to the more restrictive geometry constraints but also due to the increased sensitivity of these states with respect to interaction strength and gap-to-width ratio.}. Nevertheless, we obtain a set of four data points that satisfy all of the criteria. The topological entanglement entropy obtained from these data is $\gamma = 0.850\pm0.103$, which agrees with the Abelian theory value of $\gamma=\ln(\sqrt{5})\approx 0.805$ and iDMRG computations using the $V_1$ Haldane pseudopotential~\cite{Zaletel13}. Moreover, it is well-separated from the non-Abelian prediction of $\gamma=\ln(\sqrt{5(\varphi^2+1)})\approx1.448$, where $\varphi$ is the golden ratio~\cite{Dong08, Estienne15}. There are several additional remarks that can be made, specifically in comparison to the Laughlin state. First, the minimum $L_y/l_B$ to effectively eliminate finite-size effects is larger for the $\nu=2/5$ state than for $1/3$, increasing from 8.5 to 9.5. Second, the average spacing of the data on the $L_y/l_B$-axis is reduced. States with a larger cylinder circumference are generally more expensive to converge, and so we were not able to access high-$L_y/l_B$ states with such stringent precision. Finally, although this figure shows all of the data points obtained in accordance to the algorithm, we generally obtained a vast set of data that corroborate this conclusion, as shown in Appendix~\ref{sec:complete_data}. The Appendix also explains how one may haphazardly reach the same conclusion when not following a rigorous procedure.

The last filling factor that we consider is the $\nu=3/7$ state, where, unlike in the previous cases, the area law has not been previously investigated in any form. As before, we systematically select flux densities guided by our algorithm and we scale each configuration with $\chi$ so that we can extrapolate to the $\chi\to\infty$ limit. Due to computational expense of the $\nu=3/7$ configurations, we are not able to converge every data point to within $<0.1\%$ error and so we cannot directly use the $R^2$ value as an indicator of finite-size effects. Instead, we present the data for the largest system sizes that we examined ($L_y/l_B>10$ and $L_y=14$) with $\lesssim 3\%$ error in Fig.~\ref{fig:short_range_int}.(c). The full data set is shown in Appendix~\ref{sec:complete_data}. Interestingly, we note that there are two types of finite-size effects in the problem. Not only is there a finite-size effect due to the cylinder circumference in units of magnetic length, the physical length scale in the system, but there is also a finite-size effect due to the cylinder circumference alone, the numerical length scale. Although a large $L_y/l_B$ value ensures that each FQH droplet has a large allocated area, a large $L_y$ additionally ensures that there are enough sites (matrices in the MPS) in the finite direction of the algorithm to accurately represent the ground-state wavefunction. Since an increase in $L_y/l_B$ coarsely corresponds to an increase in $L_y$, this is an effect that was not apparent in the previous two states. In line with the observed trend, we obtain a topological entanglement entropy of $\gamma=1.020\pm0.353$, which is in agreement with the Abelian theory value of $\gamma=\ln(\sqrt{7})\approx0.973$, and well-separated from the non-Abelian prediction of $\gamma=\ln(\sqrt{7(\varphi^2+1)})\approx 1.616$~\cite{Faugno20, Dong08, Jolicoeur07, Balram19}. The derivation of the total quantum dimensions is discussed in Appendix~\ref{sec:tot_quan_dim}.

\begin{figure}
	\centering{$\nu=2/5$ state at $\kappa=1$}\\
	\includegraphics[width=\linewidth]{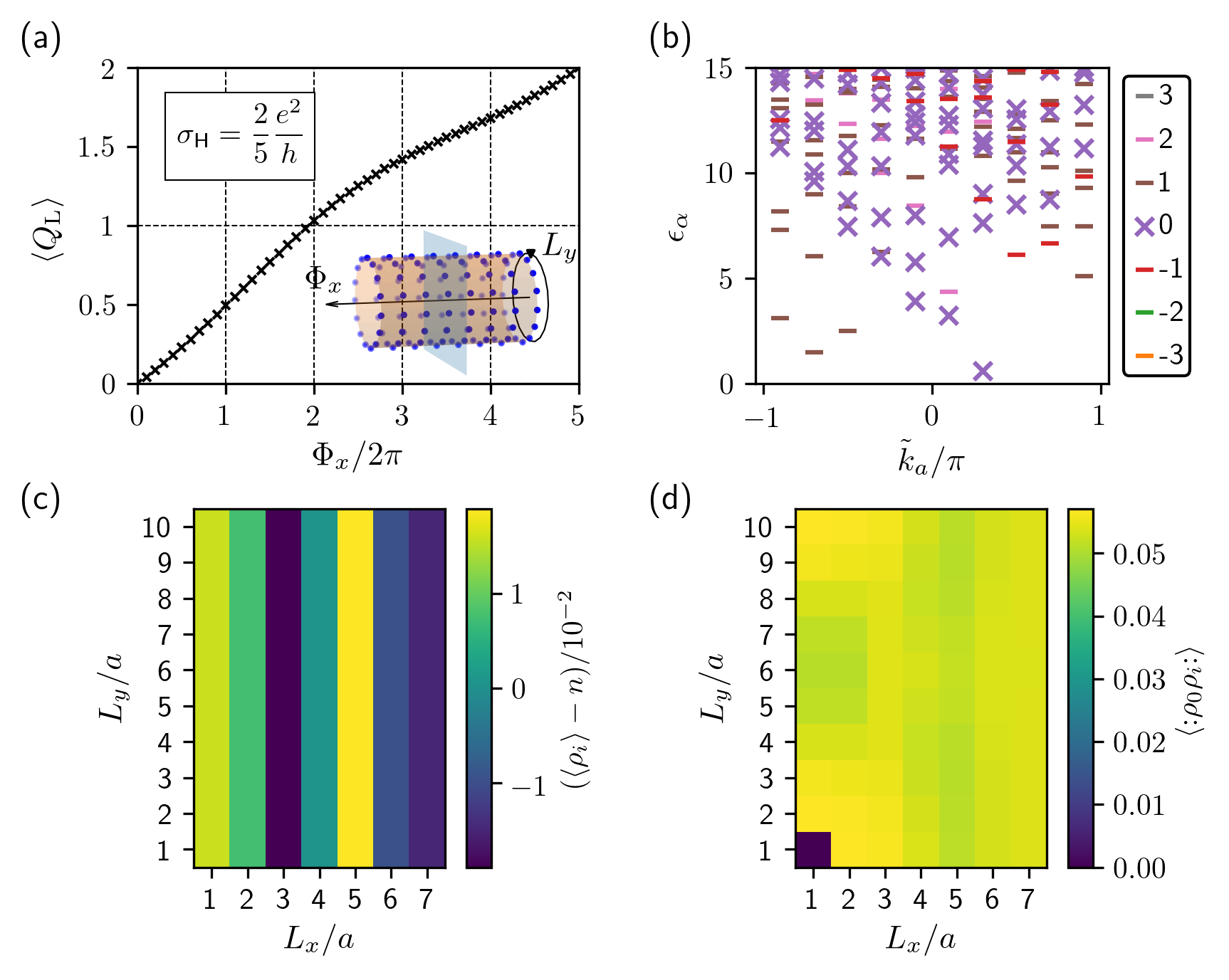}
	\centering{$\nu=3/7$ state at $\kappa=1$}\\
	\includegraphics[width=\linewidth]{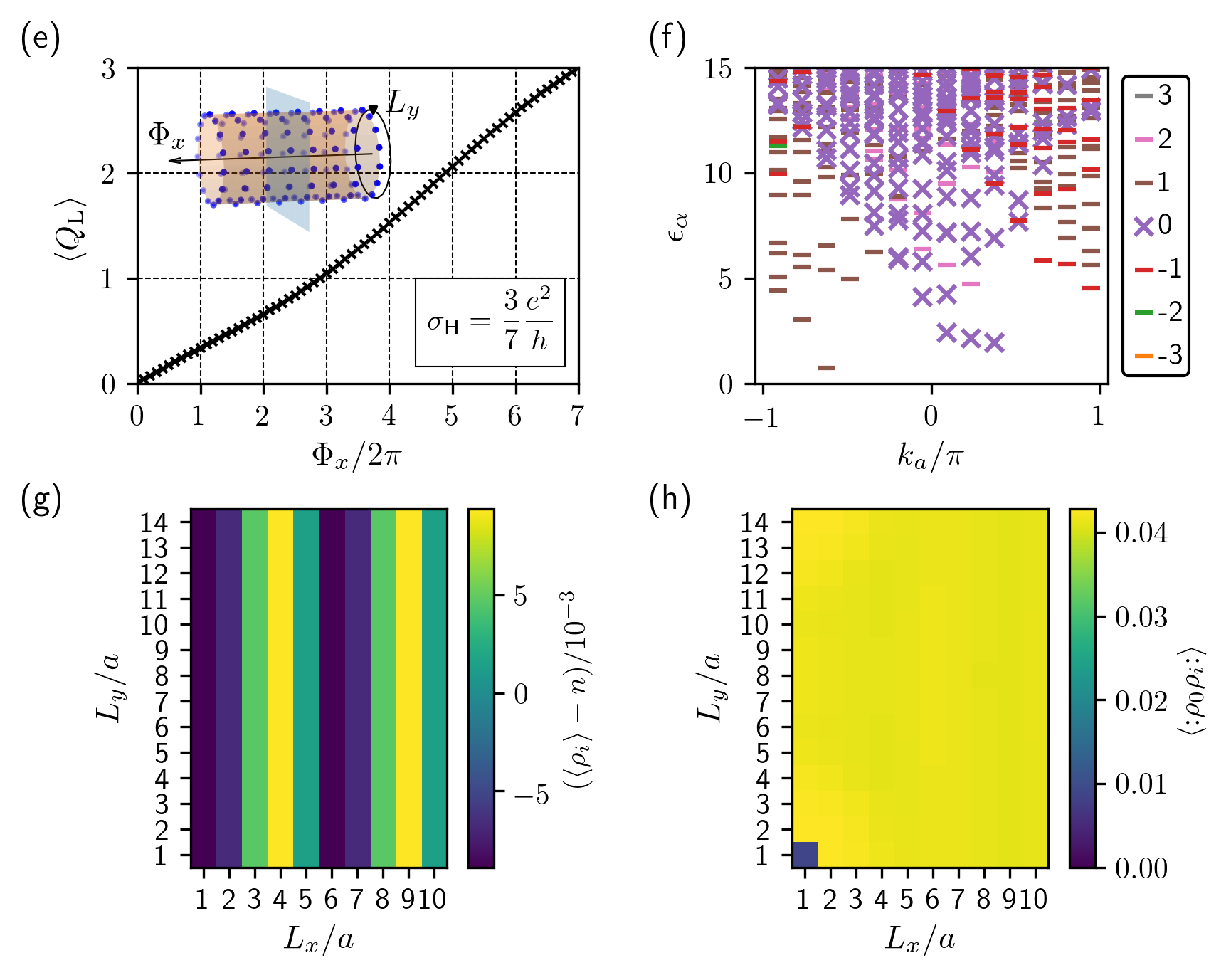}
	\caption{\label{fig:case_studies}Case studies of the points circled in red from Fig.~\ref{fig:short_range_int}. The $\nu=2/5$ state is obtained at $n_\phi=1/7$, $(L_x, L_y)=(1, 10)$, and $\chi=800$. The $\nu=3/7$ state is obtained at $n_\phi=1/10$, $(L_x, L_y)=(1, 14)$, and $\chi=2000$. (a, e) Average charge on the left half of the cylinder, $\braket{Q_\mathrm{L}}$, as a function of the external flux, $\Phi_x$, adiabatically inserted along the cylinder axis, as shown in the inset (not to scale). The charge pumping was performed at the reduced bond dimensions of $\chi=400$ and $500$, respectively. (b, f) Momentum-resolved entanglement spectrum, showing the entanglement energies, $\epsilon_\alpha$, as a function of the eigenvalues corresponding to the translation of Schmidt states around the cylinder, $k_a$, additionally colored according to their $U(1)$ charge sector. In (b) we shift the spectrum, such that $k_a\to\tilde{k}_a$, to emphasize the edge modes. (c, g) Average density, $\braket{\rho_i}$, and (d, h) two-point correlation function, $\braket{\normord{\rho_0 \rho_i}}$, for each site in the MPS unit cell. Note that the dimensions of the MPS unit cell are given in units of the lattice constant in this figure.}
\end{figure}

To confirm the FQH nature of the configurations, we additionally examine each data point in detail. In this paper, we present the case studies for the red-circled points in Fig.~\ref{fig:short_range_int}. We start by examining the details of the $\nu=2/5$ configuration at $n_\phi=1/7$ and $(L_x, L_y)=(1, 10)$, shown in Fig.~\ref{fig:case_studies}.(a-d), and subsequently the $\nu=3/7$ configuration at $n_\phi=1/10$ and $(L_x,L_y)=(1, 14)$, shown in Fig.~\ref{fig:case_studies}.(e-h). 

In Fig.~\ref{fig:case_studies}.(a), we present the charge pumping of the $\nu=2/5$ configuration as a flux is adiabatically inserted through the cylinder. The flux insertion exposes the Hall conductivity, $\sigma_\text{H}=(e^2 / h)C \nu$, where $e$ is the electronic charge, $h$ is Planck's constant, and $C$ is the Chern number. Since we are partially filling a band of unit Chern number, the Hall conductivity is directly proportional to the filling factor. The charge pumping shows that two charges are pumped across the cut after an insertion of five flux quanta, which confirms the $\nu=2/5$ FQH state. In Fig.~\ref{fig:case_studies}.(b), we present the momentum-resolved entanglement spectrum obtained by rotating the cylinder around its axis. The entanglement energies are additionally labeled by their charge sector eigenvalue, corresponding to the $U(1)$ symmetry of the Hamiltonian. Using the low-lying energies in the spectrum, we can comment on the counting of the edge states. For the $\nu=2/5$ Jain state, the counting is governed by the CFT for two non-interacting chiral bosons, which yields $1, 2, 5, 10, \dots$ with a multi-branch structure~\cite{Rodriguez13, Davenport15}. We note that we are unable to resolve the multi-branch structure in Fig.~\ref{fig:case_studies}.(b) and the verification of the counting sequence is hindered due to the modest momentum resolution, the compactified cylinder geometry, and the fact that post-Laughlin Jain states are not pure CFT states~\cite{Regnault09}. We see that the first two edge degeneracies are resolved for the bottom branch, $1, 2, \dots$, which shows similar energy gaps to the literature~\cite{Regnault09} and tentatively accords with the Abelian Jain state. However, we emphasize that the optimal configurations for the area law plot, studied in this paper, are not the optimal configurations to elucidate the edge counting. Finally, in Fig.~\ref{fig:case_studies}.(c,d) we plot the density, $\braket{\rho_i}$, and two-particle correlation function, $\braket{\normord{\rho_0\rho_i}}$, respectively. The density plot shows that we are in a striped phase and the two-point correlation function has the expected form for a conventional FQH state~\cite{Pu17, Andrews18, Schoonder19}. Most significantly, we can see from the asymmetric density and slight interference in the correlation function profile that, despite our best efforts, some minor finite-size effects still remain.  

In Fig.~\ref{fig:case_studies}.(e), we present the analogous charge pumping curve for the $\nu=3/7$ configuration. In this case, we observe three charges pumped after an insertion of seven flux quanta, again in agreement with the expected Hall conductivity for the $\nu=3/7$ FQH state in a $C=1$ band. In Fig.~\ref{fig:case_studies}.(f), we show the momentum-resolved entanglement spectrum, now at the higher resolution of $L_y=14$. In this case, we are only able to accurately resolve the first degeneracy of the bottom branch to be $1$ and hence the verification of edge-state counting is ambiguous~\cite{Rodriguez13}. Reassuringly, the density and two-point correlation function profiles in Fig.~\ref{fig:case_studies}.(g,h) show less influence of the finite system size than the $\nu=2/5$ configuration. The FQH configuration is again in a striped phase and shows the conventional correlation function profile.  

\subsection{Tuning interaction range}
\label{subsec:tune_int_range}

Having observed Abelian topological order for the $\nu=1/3$, $2/5$, and $3/7$ states when stabilized by a nearest-neighbor density-density interaction term, we now investigate the effect of increasing the interaction range, such that $1\leq\kappa\leq3$. 

\begin{figure}
	\includegraphics[width=\linewidth]{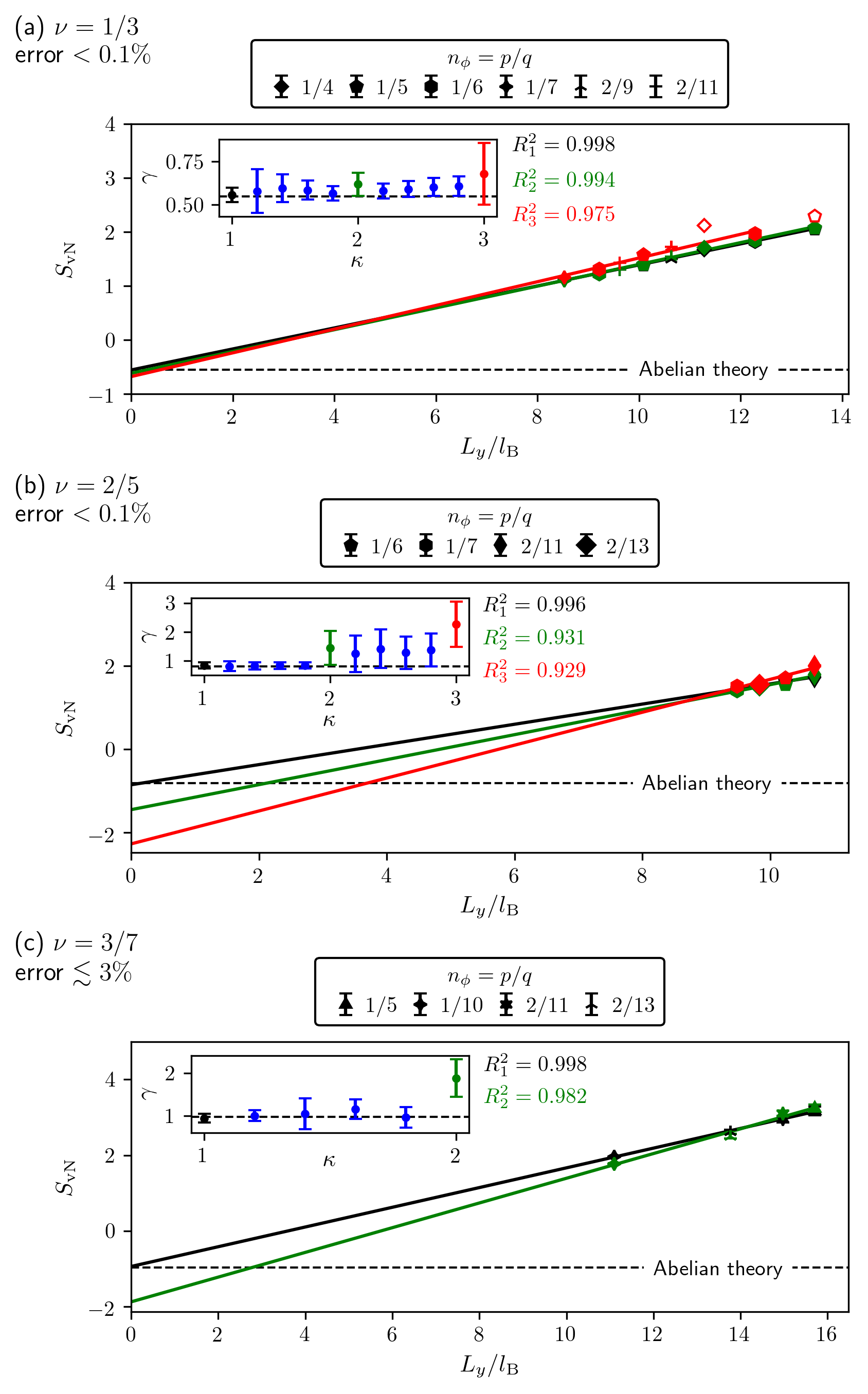}
	\caption{\label{fig:tuning_int_range}Tuning the interaction range for the area law plots in Fig.~\ref{fig:short_range_int}. In each case we tune (a, b) the corresponding colored points from Fig.~\ref{fig:short_range_int}.(a,b), and (c) the four colored points from Fig.~\ref{fig:short_range_int}.(c) that are closest to the line of best fit. The data for $\kappa=1, 2, 3$ is colored in black, green, and red, respectively, and data for intermediate values of $\kappa$ is colored blue. The value of the topological entanglement entropy predicted from the Abelian theory, $\gamma=\ln(\sqrt{s})$, is additionally marked with a dashed line. In all cases, we increase the bond dimension of these runs appropriately such that we maintain the same error threshold. Consequently, the maximum bond dimension is $\chi_\text{max}=4000$ for each filling factor. The $n_\phi=1/4$ and $1/5$, $\kappa=3$ data points at $\nu=1/3$ filling are marked as outliers.}
\end{figure}

As before, we start with the most prominent $\nu=1/3$ state. In order to investigate the effect of tuning interaction range, we take the eight accepted points in Fig.~\ref{fig:short_range_int}.(a). Subsequently, we construct equivalent area law plots for up to third nearest-neighbor interactions, as shown in Fig.~\ref{fig:tuning_int_range}.(a). At this filling factor and parameter range, our results show that there is no statistically significant increase in the topological entanglement entropy as the interaction range is increased, where the integer-$\kappa$ data yield $\gamma_{\kappa=1}=0.557\pm0.043$, $\gamma_{\kappa=2}=0.619\pm0.068$ and $\gamma_{\kappa=3}=0.680\pm0.180$. Moreover, as well as agreeing with each other with a comparable precision, all of the computed $\gamma$ agree with the Abelian theory prediction within standard error. Throughout this procedure, we ensure that the error standards are maintained to be $<0.1\%$ and the linearity threshold is consistently $R^2>0.97$. Note that the configurations at $n_\phi=1/4$ and $1/5$ with $\kappa=3$ are excluded as outliers, based on the abnormal finite-size effects observed in their density profiles. Since these configurations have the smallest system sizes and MUC dimensions ($4\times1$ and $5\times1$) out of the eight accepted points, it is unsurprising that they exhibit the largest finite-size effects for the case of up to third nearest-neighbor interactions.

In Fig.~\ref{fig:tuning_int_range}.(b) we present the analogous plot for the $\nu=2/5$ state. We take the four accepted data points from Fig.~\ref{fig:short_range_int}.(b) and maintain the error threshold to be $0.1\%$. Unlike for the $\nu=1/3$ state, at this filling factor we observe a statistically significant increase in the topological entanglement entropy as the interaction range is increased. We find that the  integer-$\kappa$ data yield $\gamma_{\kappa=1}=0.850\pm0.103$, $\gamma_{\kappa=2}=1.446\pm0.582$, and finally $\gamma_{\kappa=3}=2.267\pm0.775$. Although this statistically significant increase hinges on a couple of data points, the mean for the $\kappa \geq 2$ data is consistently and significantly larger than both the $1\leq \kappa < 2$ data and the Abelian theory prediction. More concretely, the topological entanglement entropies for $\kappa<2$ confirm the Abelian nature of the state at short interaction range and have a distinctly higher precision than the $\kappa\geq 2$ data. We caution that our result at $\kappa=3$ may be vulnerable to minor finite-size errors, particularly affecting the $n_\phi=2/11$ data point. Although we went to great lengths to alleviate finite-size effects at $\kappa=1$, the quality of the configurations is expected to deteriorate as $\kappa$ is increased. In order to overcome this, the same systematic procedure would have to be applied at $\kappa=3$, which is beyond the scope of our computational resources. The errors in the data also prevent us from commenting on the continuity of the transition in topological entanglement entropy as the interaction range is increased.

Finally, we tune the interaction range for the state at $\nu=3/7$ filling, as shown in Fig.~\ref{fig:tuning_int_range}.(c). In this case, we take the four accepted data points that are closest to the line of best fit in Fig.~\ref{fig:short_range_int}.(c). We do this in the interests of computational expense, since the configurations for this filling are already challenging to converge, even at $\kappa=1$. For consistency, we maintain the same error threshold. Following the observation for the $\nu=2/5$ state, the topological entanglement entropy increases with interaction range and by a larger margin than before, from $\gamma_{\kappa=1}=0.943\pm0.102$ to $\gamma_{\kappa=2}=1.879\pm0.441$. The $\kappa<2$ data solidifies the observation of an Abelian state at short interaction range and with a significantly higher precision than the result for $\kappa=2$. We were not able to adequately alleviate numerical and statistical errors at $\kappa=3$ for the bond dimensions that were accessible to us. However, the $\kappa=2$ data already shows a statistically significant increase from $\kappa=1$.

\begin{figure}
	\includegraphics[width=\linewidth]{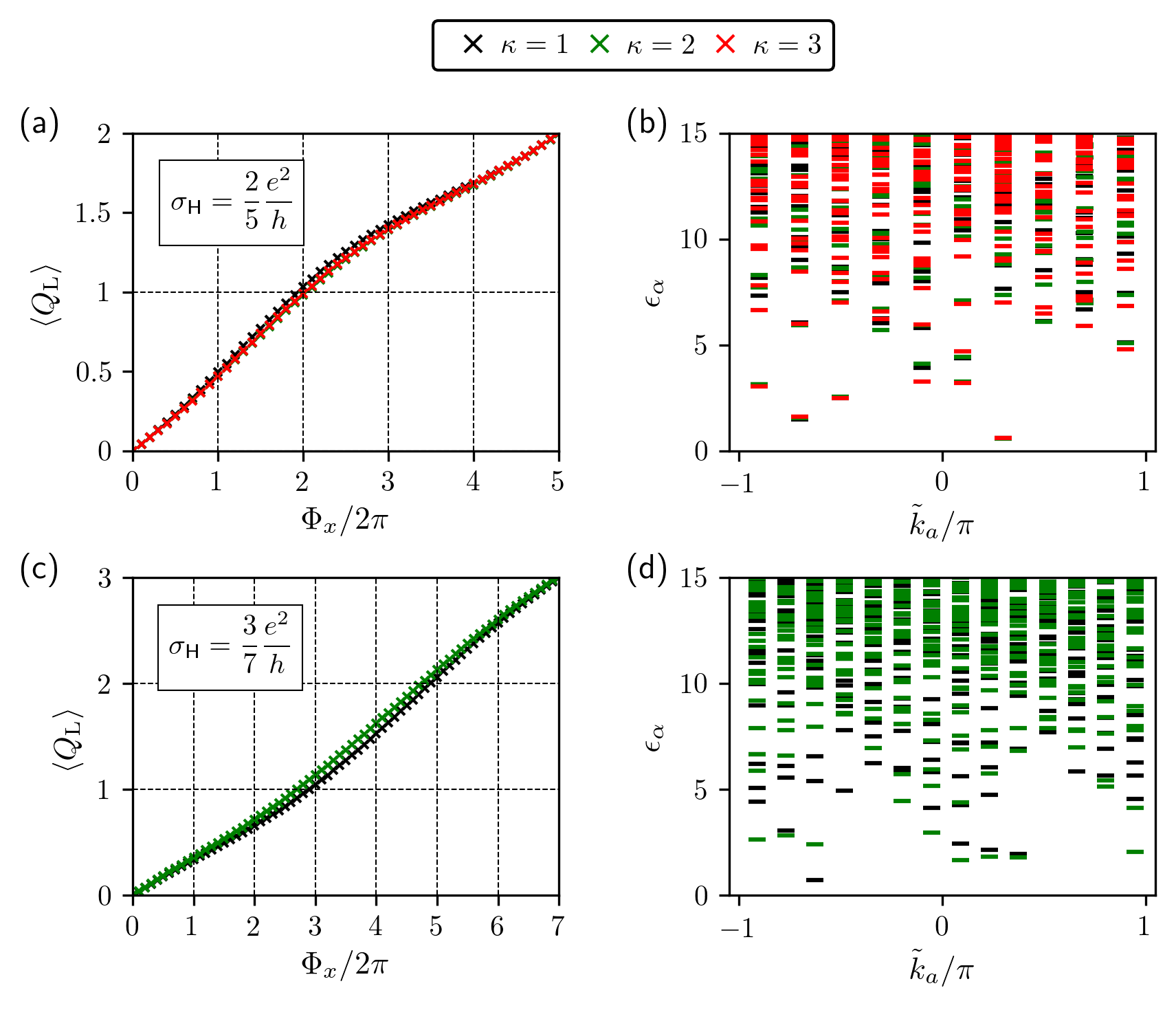}
	\caption{\label{fig:case_study_kappa}Case studies from Fig.~\ref{fig:case_studies} as a function of $\kappa$. The results for the $\nu=2/5$ state in (a, b) are obtained at $\chi=800$, $2000$, $2000$ for $\kappa=1,2,3$, respectively, whereas the results for the $\nu=3/7$ state in (c, d) are obtained at $\chi=2000$ for both $\kappa=1$ and 2. These bond dimensions correspond to the values required for the area law plot in Fig.~\ref{fig:tuning_int_range}. (a, c) Average charge on the left half of the cylinder, $\braket{Q_\mathrm{L}}$, as a function of the external flux, $\Phi_x$, adiabatically inserted along the cylinder axis. The charge pumping was performed at the reduced bond dimensions of $\chi=400$ for $\nu=2/5$ at $\kappa=1$, and $\chi=500$ for all other states. (b, d) Momentum-resolved entanglement spectra, showing the entanglement energies, $\epsilon_\alpha$, as a function of the eigenvalues corresponding to the translation of Schmidt states around the cylinder, $k_a$. Where necessary, we shift the spectra to emphasize the edge modes.}
\end{figure}

In order to verify the robustness of the FQH states at $\kappa>1$, we examine their charge pumping and momentum-resolved entanglement spectra. In this paper, we present in Fig.~\ref{fig:case_study_kappa} the corresponding plots for the case studies introduced in Fig.~\ref{fig:case_studies}. From the charge pumping curves in Fig.~\ref{fig:case_study_kappa}.(a,c), we can confirm that the higher-$\kappa$ states are valid FQH states corresponding to $\nu=2/5$ and $3/7$ filling of a $|C|=1$ band. From the $\nu=2/5$ entanglement spectra in Fig.~\ref{fig:case_study_kappa}.(b), we notice only minor deviations in the low-lying entanglement energies, which become more pronounced as $\kappa$ increases from 2 to 3. From the $\nu=3/7$ entanglement spectra in Fig.~\ref{fig:case_study_kappa}.(d), we observe a significant deviation in the low-lying states as $\kappa$ is increased, where only a small minority of the $\kappa=1$ and $2$ energies overlap. In both cases, the numerical configurations preclude a precise verification of the counting.

\subsection{Tuning interaction strength}
\label{subsec:tune_int_strength}

Having observed an increase in the topological entanglement entropy with interaction range, we now investigate the effect of increasing the interaction strength, such that $10\leq V_0 \leq 50$, at fixed interaction range $\kappa=1$.

\begin{figure}
	\includegraphics[width=\linewidth]{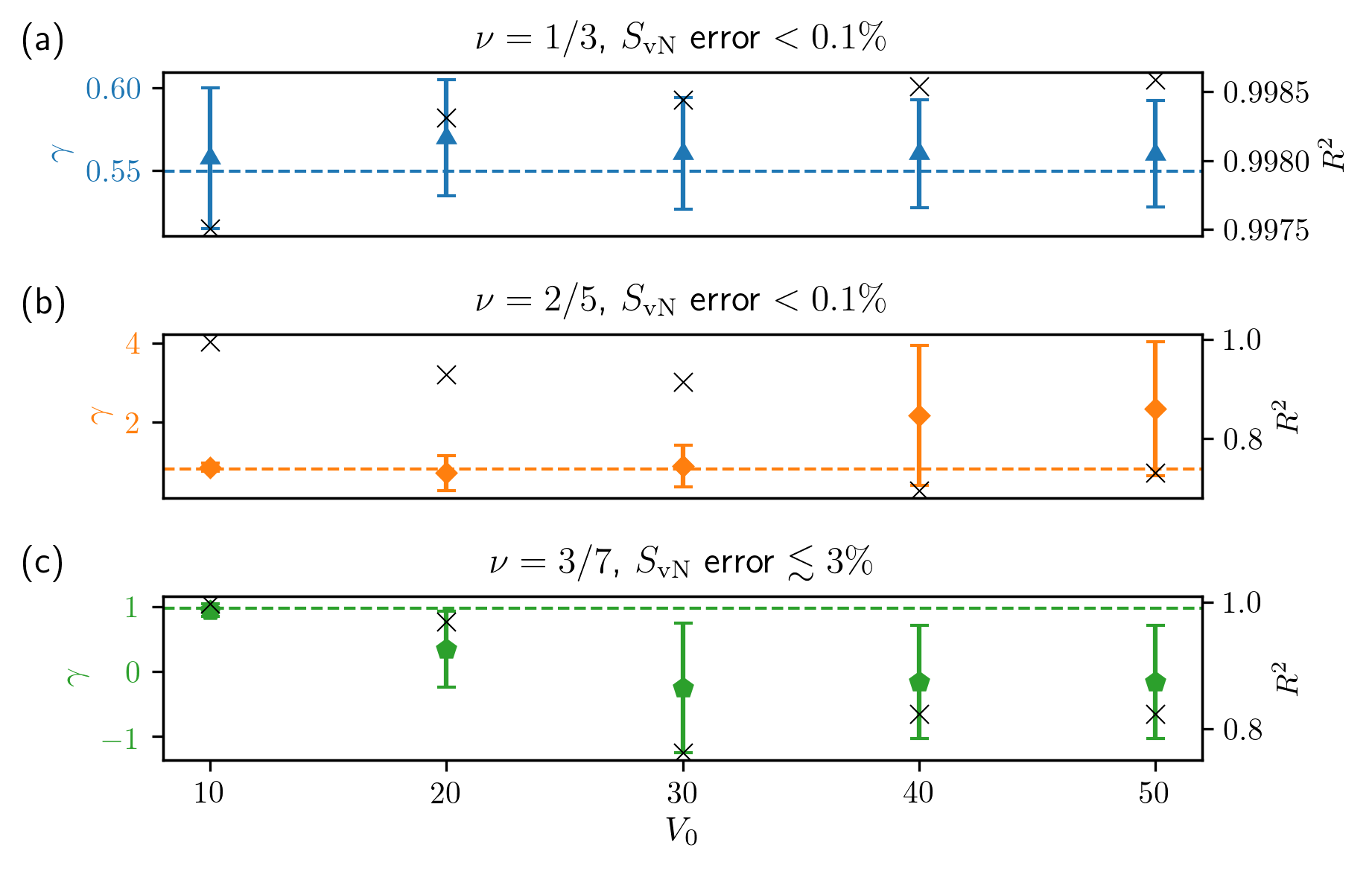}
	\caption{\label{fig:tuning_int_strength}Tuning the interaction strength for the $\kappa=1$ area law plots in Fig.~\ref{fig:tuning_int_range}. The topological entanglement entropy data for the (a)~$\nu=1/3$, (b)~$2/5$, and (c)~$3/7$ series are colored blue, orange, and green, respectively. On the same axes, we plot the $R^2$ values of the corresponding area law plots with black crosses. The value of the topological entanglement entropy predicted from the Abelian theory, $\gamma=\ln(\sqrt{s})$, is marked with a dashed line. In all cases, the bond dimensions are the same as those for the $\kappa=1$ plots in Fig.~\ref{fig:tuning_int_range}.}
\end{figure}

In Fig.~\ref{fig:tuning_int_strength}, we present the topological entanglement entropies for the $\nu=1/3$, $2/5$, $3/7$ fillings, along with their associated $R^2$ values, as we vary $V_0$ for the $\kappa=1$ configurations in Fig.~\ref{fig:tuning_int_range}. In Fig.~\ref{fig:tuning_int_strength}.(a), we see that the result for $\nu=1/3$ is stable to increases in interaction strength. In fact, the error in the topological entanglement entropy decreases from $\gamma_{V_0=10}=0.557\pm0.043$ to $\gamma_{V_0=50}=0.560\pm0.032$ and the $R^2$ value monotonically increases from $R^2_{V_0=10}=0.998$ to $R^2_{V_0=50}=0.999$, which indicates that the result is converging, albeit slightly. In contrast, we see that in Fig.~\ref{fig:tuning_int_strength}.(b), the result for $\nu=2/5$ is not robust to increases in the interaction strength. At $V_0=20$ and $30$, the $n_\phi=1/6$ configuration deviates from the linear correlation, which negatively impacts the $R^2$ values and error bars. Moreover, at $V_0=40$ and $50$, two of the four configurations deviate from the linear correlation, which renders an extrapolation of the topological entanglement entropy meaningless. In Fig.~\ref{fig:tuning_int_strength}.(c) we observe a breakdown of the extrapolation for the $\nu=3/7$ state already at $V_0=20$. In this case, the topological entanglement entropy fluctuates significantly for $V_0\geq20$, which is a clear indication of numerical error\footnote{The topological entanglement entropy is also negative, which is not physical.}. This analysis shows that the selected FQH configurations for the $\nu=2/5$ and $3/7$ states are prohibitively sensitive to increases in the interaction strength in the parameter range we study. In order to successfully compute the topological entanglement entropies for these states, one would have to reevaluate the algorithm for each value of $V_0$, which is currently beyond the scope of our computational resources.  

\section{Discussion and Conclusions}
\label{sec:discussion}

In this paper, we have investigated the Abelian topological order for the single-component $\nu=1/3$, $2/5$, and $3/7$ FQH states in the Hofstadter model with band mixing. Having developed an efficient sampling algorithm that accounts for both numerical and statistical errors, we constructed the area law of entanglement for each of these states and extrapolated to read off the topological entanglement entropies. For all states, we demonstrated Abelian topological order when interactions are nearest-neighbor. Subsequently, we investigated the effect of increasing the interaction range and strength. The non-Laughlin FQH configurations are sensitive to increases in interaction range, where we observe a corresponding increase in topological entanglement entropy, and interaction strength, to the extent that we can no longer reliably extrapolate to the thermodynamic limit. In contrast, the extrapolation for the Laughlin state is robust in both cases. These results highlight the sensitivity of Abelian FQH states in lattice models, as well as the scope of our proposed algorithm.

Given the consistent Abelian result for the topological entanglement entropy in the case of nearest-neighbor interactions, we comment briefly on the nature of these states as the interaction range and strength are increased. 

In Fig.~\ref{fig:tuning_int_range}, we observe an increase in topological entanglement entropy with $\kappa$ for the $\nu=2/5$ and $3/7$ states. There are several possible explanations for this increase, including: (i) a breakdown of the FQH state, (ii) numerical error, or (iii) a transition to different quantum statistics. Scenario (i) postulates a transition from the FQH regime to a competing phase, which is conceivable, particularly in the band mixing regime~\cite{Zhao18}. However, by studying the charge pumping and entanglement spectra in Fig.~\ref{fig:case_study_kappa}, it is evident that the FQH phases have not broken down. Scenario (ii) postulates that finite-size effects become uncontrolled as the interaction range is increased, which makes the data prohibitively noisy. Although the data is more susceptible to these effects at large $\kappa$, we place a strong emphasis on error analysis and only use the highest-quality subset of data points in Fig.~\ref{fig:tuning_int_range}. Moreover, if numerical errors were dominant, one would expect fluctuations in the topological entanglement entropy, rather than the steady increase that we observe. Scenario (iii) postulates that the Abelian FQH states transition to different quantum statistics, such as non-Abelian order~\cite{Liu13_2}. With respect to the topological entanglement entropy, our data do not rule this out. Although the $\kappa=3$ data at $\nu=2/5$ slightly exceeds the non-Abelian prediction of $\gamma\approx1.448$, this may be attributed to the minor finite-size effects present for the $n_\phi=2/11$ data point.

In Fig.~\ref{fig:tuning_int_strength}, we observe that the original extrapolation of the topological entanglement entropy for the $\nu=1/3$ state is stable to increases in interaction strength, whereas the extrapolations for the $\nu=2/5$ and $3/7$ states are not. Indeed, the Laughlin state is known to be more robust than higher-order FQH states and has been shown numerically on a lattice to survive with interaction strengths that far exceed the band gap~\cite{Kourtis14}. Moreover, we obtained a comparatively vast data set for the $\nu=1/3$ state, which allowed us to tighten the constraints of the algorithm. As a result, the error of the entropy values for the $\nu=1/3$ data is at least an order of magnitude smaller than the $\nu=2/5$ state and two orders of magnitude smaller than the $3/7$ state. The finite-size effects are also reduced for the $\nu=1/3$ state in a comparable $L_y/l_B$ domain, which renders the topological entanglement entropy significantly more robust. For the $\nu=2/5$ and $3/7$ states, we notice that the increased interaction strength affects the noise in our data to the extent that we can no longer reliably extrapolate to the thermodynamic limit. To accurately decouple the physical and numerical differences between the Laughlin and non-Laughlin states and comment on the topological entanglement entropy, the algorithm would need to be reevaluated for larger values of interaction strength.            

This work complements several recent investigations in both experiment and theory. In the continuum, there is a plethora of experimental phenomenology to show that the Jain series of states from $\nu=1/3$ to $1/2$ are Abelian for the long-range Coulomb interaction~\cite{Jain07}. In particular, the first direct experimental observation of fractional statistics was made by two independent groups last year for the $\nu=1/3$ state, which confirm its Abelian nature~\cite{Nakamura20, Bartolomei20}. Moreover, theoretical investigations have been made into non-Abelian counterparts for the Jain series~\cite{Bernevig08}. We note, however, that apart from the Read-Rezayi clustered states with many-body interactions~\cite{Read99}, these states are derived from non-unitary conformal field theories, which casts doubt on whether they can describe gapped topological phases~\cite{Read09}. There have also been iDMRG investigations of the topological entanglement entropy of the $\nu=2/5$ state in the continuum~\cite{Zaletel13}, as well as equivalent computations for trial wavefunctions directly transcribed to a MPS representation~\cite{Estienne15}, which all take the Abelian value. In lattice models, there has been significant progress in experiments using optical flux lattices~\cite{Goldman16} and twisted bilayer graphene~\cite{Spanton18}. In addition, numerical investigations have found that the interaction range and strength can increase the stability of FQH states in optical lattices~\cite{Hafezi07}, and they have identified numerous experimental proposals to realize lattice FQH states in bilayer graphene~\cite{Ledwith20, Repellin20}. On the other hand, although theory suggests that the statistics of elementary excitations for lattice FQH states may be sensitive to interaction range~\cite{Liu13_2}, experimental investigations in this direction for non-Laughlin lattice FQH states are still limited.

The results in this paper extend this foundation in three ways. First, we investigate the $\nu=2/5$ and $3/7$ states using iDMRG in lattice models. This is in contrast to prior research, where the topological entanglement entropy has only been investigated using iDMRG for the $\nu=1/3$ state in lattice models~\cite{Schoonder19, Andrews20} and the $\nu=2/5$ state in the continuum~\cite{Zaletel13}. Second, in order to investigate the lattice $\nu=2/5$ and $3/7$ states, we overcame computational and statistical challenges by developing an algorithm to construct the area law plot. Finally, we exploited this algorithm to tune the interaction range and strength, and we commented on the scope of the algorithm and nature of the states in these regimes. Finding effective ways to determine the topological order hosted by such prominent FQH plateaus is important, not only to bolster our understanding of lattice FQH states but also due to its practical implications. For example, some manifestations of the lattice FQH states discussed in this paper have already been observed experimentally in bilayer graphene~\cite{Faugno20}. Moreover, advancements in optical lattices offer a promising way to tune the interaction parameters for custom lattice configurations~\cite{Cooper19}. 

Future studies in this area could seek to characterize the states at large $\kappa$ and $V_0$. Specifically, the complete anyonic statistics have not yet been obtained from a comparable many-body simulation of the $\nu=2/5$ or $3/7$ state in these limits. Other avenues for research include numerical simulations on the torus to confirm the incompressibility of the $\nu=2/5$ and $3/7$ states~\cite{Jolicoeur14}, as well as work to establish the prerequisites for stabilizing exotic Abelian states (with $\gamma>\ln(\sqrt{s})$). On the technical side, it would be interesting to compare our results with the approach developed by Zaletel et al., which implements the Coulomb interaction using Haldane pseudopotentials on the infinite cylinder~\cite{Zaletel13, Zaletel15}. Coupled with this, it would be instructive to examine the effect of a dipolar interaction, to see whether the effect on topological entanglement entropy is accentuated.

We hope that this paper will not only emphasize the care required when constructing area laws of entanglement but, more broadly, highlight the sensitivity of Abelian lattice FQH states with respect to both interaction range and strength. 

\begin{acknowledgments}
We thank Ajit Balram, Johannes Hauschild, Apoorv Tiwari, William Faugno, Leon Schoonderwoerd, and Gunnar M{\"o}ller for useful discussions. In particular, we thank Ajit Balram for sharing the latest version of his preprint prior to publication~\cite{Faugno20}. Calculations were performed using the \textsc{TeNPy} Library (version 0.5.0)\cite{tenpy} and \textsc{GNU Parallel}~\cite{Tange11}. This project was funded by the Swiss National Science Foundation under Grant No.~PP00P2\_176877.
\end{acknowledgments}

\appendix

\section{Numerical method}
\label{sec:num_method}

In an effort to remove arbitrariness from the computation of topological entanglement entropies, we follow a set algorithm for the data collection and analysis. In this appendix, we outline and justify the steps in the procedure. We start with a description of the systematic selection of $n_\phi$ values in Sec.~\ref{subsec:nphi}, we then explain how the entanglement entropy is extrapolated for each configuration in Sec.~\ref{subsec:extrapolation}, and finally show how the line of best fit is constructed for the area law of entanglement in Sec.~\ref{subsec:line}. 

\subsection{Selection of $n_\phi$}
\label{subsec:nphi}

For a given filling factor, we list all values of the coprime fraction $n_\phi\equiv p/q$ that satisfy the following constraints:

\begin{itemize}
	\item $\displaystyle \frac{1}{2\pi}\left( \frac{(L_y/l_B)_\text{min}}{L_y} \right)^2 < n_\phi < 0.4$
	
	Due to finite-size effects, we can safely set a lower bound on the $L_y/l_B = \sqrt{2\pi n_\phi}L_y$ values that we produce. From preliminary investigations of the bosonic Laughlin state [Fig.~\ref{fig:complete_data}.(a)], the simplest FQH state, we found that finite-size effects are sufficiently suppressed at $L_y/l_B \gtrsim 8$. Since finite-size effects for higher-order FQH states require larger cylinder circumferences to be suppressed, we set at least $(L_y/l_B)_\text{min}\equiv8$. As we proceed up the FQH hierarchy, we can increase this minimum threshold. Furthermore, it has been shown experimentally that Laughlin states require $n_\phi<0.4$ to be stabilized and that this critical $n_\phi$ may be lower for higher-order hierarchy states~\cite{Hafezi07}. Hence, we set the upper limit for the flux densities to $0.4$.  

	\item $\displaystyle L_y \geq 4$
	
	Since we are interested in the effect of tuning the interaction range across first, second, and third nearest neighbors, we demand that the system size is at least four sites across in the $y$-direction.
	
	\item $\displaystyle 4\leq q \leq 20$
	
	Similar to above, we demand that the system size in the $x$-direction is also at least four sites across\footnote{These conditions on $L_y$ and $q$ do not guarantee that a system will be large enough to sufficiently suppress finite-size effects, as exemplified by the outliers in Fig.~\ref{fig:tuning_int_range}.(a).}. Furthermore, we set the maximum value to $20$ so that we limit the precision of the required flux density values. This makes the results more relevant for experimental set-ups, such as optical flux lattices~\cite{Jaksch03}. 
	
	\item $\displaystyle N_\text{min}\geq 2$
	
	We require that the total number of particles in our system ($N=n q L_x L_y$) is an integer greater than or equal to two, for two-body interactions.
	
	\item $\displaystyle q L_x L_y \leq N_{\text{s},\text{max}}$
	
	We limit the total size of the MPS unit cell to be less than or equal to $N_{\text{s},\text{max}}$. This restricts the memory cost, which scales linearly with the system size. We adjust the value $N_{\text{s},\text{max}}\sim100$ depending on computational resources.
	
	\item $\displaystyle \Delta_{L_y/l_B}>0.1$
	
	We require that the separation between $L_y/l_B$ values is greater than 0.1. This is to minimize the susceptibility of the line of best fit to errors in the individual values for the entanglement entropies. It also efficiently provides a greater range of $L_y/l_B$ values to consider. The value of $0.1$ was chosen as a compromise between precision and a maximal data set.
	
\end{itemize}

Finally, we define a processing cost function $\Gamma(p,q,L_x,L_y)\equiv qL_x \mathrm{e}^{L_y+\tilde{n}_\phi}$, where $\tilde{n}_\phi = 20 n_\phi$ is the normalized flux density, defined such that the range $1\leq \tilde{n}_\phi < 8$ is comparable to $6\lesssim L_y \lesssim 15$. We then sort the configurations in ascending $\Gamma$, attempt to converge all of them up to $\chi_\text{max}\lesssim3000$, and use those that are successful.

The processing cost function roughly quantifies how much processing time is needed for a state to converge. It is not the physical processing cost of the iDMRG algorithm itself, which scales (conservatively) as $\sim O(\chi^3 D d^3 + \chi^2 D^2 d^2)$ for a single bond update, where $D$ is the MPO bond dimension and $d$ is the single-site Hilbert space dimension. Rather, the processing cost function takes into account the $\chi$ required for convergence in a given model. It is well known that the convergence processing cost of the iDMRG algorithm scales linearly with cylinder length and exponentially with cylinder circumference. Moreover, from preliminary investigations of the Laughlin states, we find that a state is more likely to converge with a smaller $n_\phi$, showing roughly the same scaling precedence as cylinder circumference.

\subsection{Extrapolation of the entanglement entropy to the $\chi\to\infty$ limit}
\label{subsec:extrapolation}

\begin{figure*}
	\includegraphics[width=\linewidth]{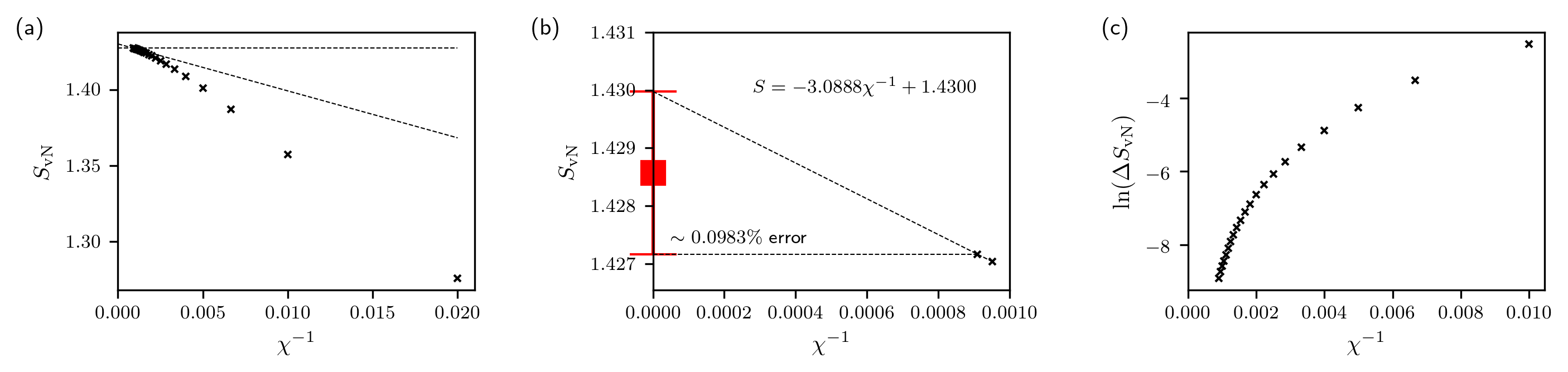}
	\caption{\label{fig:entropy_conv}Extrapolation of the entanglement entropy, $S_\mathrm{vN}$, in the $\chi\to\infty$ limit for the fermionic Hofstadter model at $\nu=1/3$, with flux density $n_\phi=1/3$ and cylinder circumference $L_y=6$. (a) $S_\mathrm{vN}$ as a function of $\chi^{-1}$, where $\chi\in[50, 100, 150, \dots , 1100]$. (b) A close-up of the last two points shown in (a), including the equation of the straight line through these points, and the $\lim_{\chi\to\infty}(S_\mathrm{vN})$ estimate with error bars shown in red. (c) The change in entropy as the bond dimension is incremented, $\Delta S_\mathrm{vN}$, for the data points in (a).}
\end{figure*}

Since the final result for the topological entanglement entropy, $\gamma$, is highly sensitive to the values of the individual entanglement entropies, $S_\mathrm{vN}$, we need to minimize the error in $S_\mathrm{vN}$ to obtain a representative value for $\gamma$. Furthermore, since we are comparing $\gamma$ between different Hamiltonians, we additionally need to ensure that all $S_\mathrm{vN}$ are computed to the same accuracy for a fair comparison.

To this end, we study the convergence of $S_\mathrm{vN}$ with bond dimension $\chi$. An illustrative example for the Laughlin state in the fermionic Hofstadter model is shown in Fig.~\ref{fig:entropy_conv}. Figures~\ref{fig:entropy_conv}.(a,b) show the convergence of $S_\mathrm{vN}$ with $\chi$ for this system. We can see that as $\chi$ is increased, the entropy is approaching a value of $1.43$. To quantify this, we note that entropy increases monotonically with bond dimension, as demonstrated in Fig.~\ref{fig:entropy_conv}.(c). Hence, the highest-$\chi$ value for $S_\mathrm{vN}$ will be the lower bound of our entropy estimate. Moreover, we know that in the $\chi\to\infty$ limit, $\mathrm{d}S_\mathrm{vN}/\mathrm{d}\chi\to0$, which implies that $\mathrm{d}S_\mathrm{vN}/\mathrm{d}\chi$ monotonically decreases. Hence the extrapolation of our last estimate of $\Delta S_\mathrm{vN}/\Delta\chi$ will serve as our upper bound. Since a polynomial fit of all data points is computationally costly, and will have negligible benefits as we reduce the errors, we instead take $\lim_{\chi\to\infty}S_\mathrm{vN}$ to be directly in between our lower and upper bounds, as exemplified by the red data point in Fig.~\ref{fig:entropy_conv}.(b).

In order to maintain a consistent accuracy among all entanglement entropy values in this manuscript, we continue with the entropy convergence until all $\lim_{\chi\to\infty}S_\mathrm{vN}$ estimates have error bars $<0.1\%$, unless otherwise stated. Consequently, the example system in Fig.~\ref{fig:entropy_conv} is sufficiently converged for $\chi_\text{max}=1100$.

\subsection{Linear regression}
\label{subsec:line}

For a given area law plot of the entanglement entropy, there can be great variability in the $y$-intercept of the linear regression depending on which points are considered, as shown in Fig.~\ref{fig:complete_data}. Moreover, it is known that finite-size effects become significant for small systems, and in the extreme case, the area law even breaks down since $S=\alpha L_y - \gamma + O(\mathrm{e}^{-L_y})$. Therefore, there is motivation to carefully reject data with small cylinder circumferences without biasing the final result for the topological entanglement entropy. To reconcile this issue, we use an algorithm to construct the line of best fit.

For all the data points on the plot, we draw lines of best fit: the first of which considers all of the data, the second rejects the smallest $L_y/l_B$ point, the third rejects the smallest two $L_y/l_B$ points, etc. We continue in this manner until we reach a line that satisfies $R^2>0.99$. It is this line that we use for our linear regression. The data is quantifiably linear and so finite-size effects are suppressed, and we take the first such line because it is based on the most points. 

\section{Complete data sets}
\label{sec:complete_data}

\begin{figure*}
	\includegraphics[width=\linewidth]{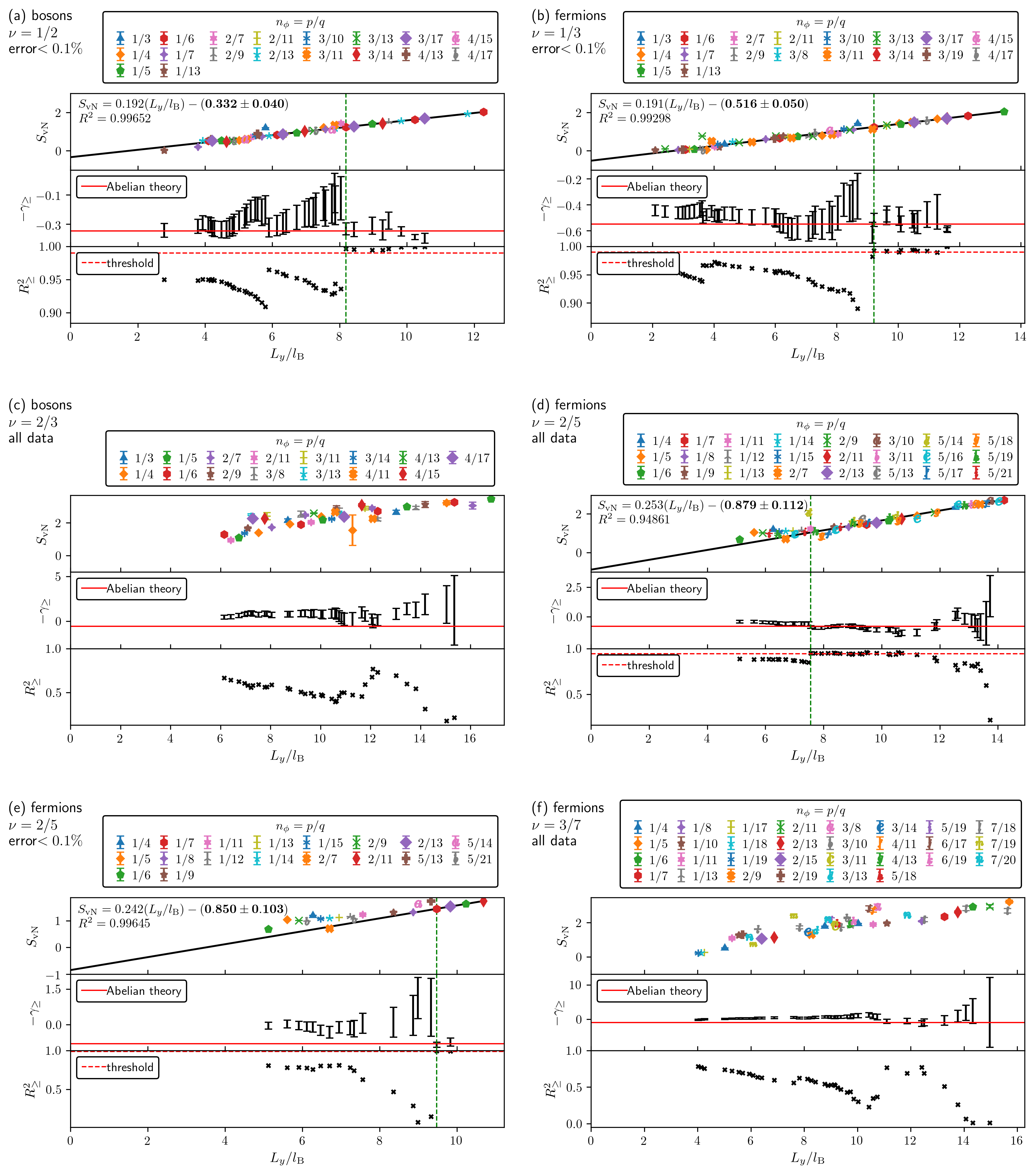}
	\caption{\label{fig:complete_data}The complete data sets for the: (a) bosonic and (b) fermionic Laughlin (0th hierarchy) states with $<0.1\%$ error; the (c) complete bosonic, (d) complete fermionic, and (e) $<0.1\%$ error fermionic 1st hierarchy states; and (f) the fermionic 2nd hierarchy state. (top panels) Von Neumann entanglement entropy, $S_\text{vN}$, plotted as a function of cylinder circumference, $L_y$, in units of magnetic length, $l_B=(2\pi n_\phi)^{-1/2}$. The line of best fit is drawn through the set of points above the smallest $L_y/l_B$ that yields $R^2>0.99$ for $<0.1\%$ error data and $R^2=R_\text{max}^2$ otherwise. This cut-off is marked with a green dashed line. (middle panels) The $y$-intercept of the line of best fit drawn through all points greater than or equal to $L_y/l_B$, denoted as $-\gamma_{\geq}$. The Abelian theory prediction, $\gamma=\ln(\sqrt{s})$, is marked in red. (bottom panels) The square of Pearson's correlation coefficient for the line of best fit drawn through all points greater than of equal to $L_y/l_B$, denoted as $R_{\geq}^2$. The corresponding threshold, either $R^2=0.99$ or $R_\text{max}^2$, is marked with a red dashed line.}
\end{figure*}

In Fig.~\ref{fig:complete_data}, we present the complete set of data collected in this project for $\kappa=1$, both systematic and unsystematic. Along with the area law of entanglement in the top panel of each plot, we also present the topological entanglement entropy estimate from including all points with a cylinder circumference $\geq L_y/l_B$ (middle panel), along with the corresponding $R^2$ values for each of these linear regressions (bottom panel). 

In Fig.~\ref{fig:complete_data}.(a,b) we show data for the bosonic and fermionic Laughlin states. For the bosonic Laughlin state, we computed the entanglement entropy for 76 configurations, 46 of which converged with $<0.1\%$ error and are shown in Fig.~\ref{fig:complete_data}.(a). Even with the complete $<0.1\%$ error data set, we still obtain a clear agreement with the Abelian theory value of $\gamma=\ln(\sqrt{2})\approx0.347$. We also note that after the $R^2=0.99$ threshold, finite-size effects are suppressed since we maintain $R^2>0.99$, and the average topological entanglement entropy agrees with the theory based on all values drawn after this point. This plot also highlights the dramatic effect that a few relatively minor outliers (notably $\{n_\phi=1/3$, $L_y=4\}$ and $\{n_\phi=2/7$, $L_y=6\}$) can have on the topological entanglement entropy estimate and raises concerns of selection bias in unsystematic studies. For the fermionic Laughlin state, we computed the entanglement entropy for 89 configurations, 53 of which converged and are shown in Fig.~\ref{fig:complete_data}.(b). Overall, we notice similar features as for the bosonic Laughlin state. However, we note that on this occasion the complete $<0.1\%$ error data set does not agree with the Abelian theory. We emphasize that, as with the bosonic Laughlin state, it is simply coincidence whether or not the complete data set agrees with the theory due to the significant finite-size effects at small cylinder circumference. Once we draw a line through all points above the $R^2=0.99$ threshold, we do see a clear agreement. As with the bosonic Laughlin state, $R^2>0.99$ is maintained above this threshold, which indicates that finite-size effects have been effectively alleviated. Note also that the $R^2>0.99$ threshold occurs at $L_y/l_B=9.21$, which is larger than the bosonic value $L_y/l_B=8.12$, as expected.
 
Motivated by the results from the Laughlin states, we construct corresponding plots for the next filling factors in the hierarchy: the bosonic $\nu=2/3$ and fermionic $\nu=2/5$ states, shown in Fig.~\ref{fig:complete_data}.(c--e). For the bosonic $\nu=2/3$ state, we computed the entanglement entropy for 41 configurations, only 5 of which converged, and so we simply show all of the data in Fig.~\ref{fig:complete_data}.(c). For this filling factor, we are not able to draw any conclusions regarding the topological entanglement entropy. We note incidentally, however, that the value of the topological entanglement entropy for the largest $R^2$ value gives the closest agreement with the Abelian theory, which indicates potential agreement once finite-size effects are alleviated. For the fermionic $\nu=2/5$ state we computed 50 entanglement entropies, shown in Fig.~\ref{fig:complete_data}.(d), and 20 converged to $<0.1\%$ error, shown in Fig.~\ref{fig:complete_data}.(e). In this case, when a line is drawn through \emph{all} data points above the $R_\text{max}^2$ threshold in Fig.~\ref{fig:complete_data}.(d), we obtain a close agreement to the Abelian theory value, similar to the result obtained using the $<0.1\%$ error data. We note however, that the $R^2$ value does not maintain its large value after the threshold, which indicates that significant fluctuations are still present. In contrast, when we examine exclusively $<0.1\%$ error data points in Fig.~\ref{fig:complete_data}.(e), we see that $R^2>0.99$ is maintained after the threshold. In keeping with the noted trend, the value at which this occurs, $L_y/l_B=9.47$, is higher than for the corresponding Laughlin state.  

Most ambitiously, we construct an area law plot for the secondary fermionic hierarchy state at $\nu=3/7$. For this state, we computed the entanglement entropy for 44 configurations, only 12 of which converged to $<0.1\%$ error. Consequently, we present all of the data in Fig.~\ref{fig:complete_data}.(f). Given the immense computational effort in obtaining high-$L_y/l_B$ data for the $\nu=3/7$ state ($\sim256$GB of memory and $\sim2$ weeks run-time per data point), we analyze this as a stand-alone plot. As demonstrated in Fig.~\ref{fig:complete_data}.(d), we cannot reliably apply the $R^2$ threshold analysis, since the error of the data set is not small enough. We also note that for this system both physical ($L_y/l_B$) and numerical ($L_y$) finite-size effects are significant. Taking the largest system sizes that we considered, with respect to both length scales, yields the subset shown in Fig.~\ref{fig:short_range_int}.(c). The extracted data is still significantly noisier than any other data presented in the main text, however the complete data set suggests that these finite-size effects will be alleviated if the cylinder circumference is further increased. Specifically, we again point out that the estimate of the topological entanglement entropy with the largest $R^2$ value yields the closest agreement to the Abelian theory. This is true for all of the area law plots in our analysis.     

\section{Total quantum dimensions}
\label{sec:tot_quan_dim}

Generalized parafermion FQH states are expected at the filling $\nu=k/(Mk+2)$, where $M$ even/odd corresponds to bosons/fermions and $k$ is an integer~\cite{Read99}. These states are described by a $SU(2)_k$ Chern-Simons theory in the bulk and the rational CFT $[SU(2)/U(1)]_k \times U(1)_{k(MK+2)}$ on the edge~\cite{Dong08}. Since the total quantum dimensions for $SU(2)$ or $U(1)$ theories with positive and negative $k$ levels are identical, the total quantum dimension for the coset theory is given as
\begin{equation}
\mathcal{D}_\text{coset}=\frac{\sqrt{(|k|+2)|Mk+2|}}{2\sin\left(\frac{\pi}{|k|+2}\right)}. 
\end{equation}
For the non-Abelian $\nu=2/5$ state we may set $k=3$, $M=1$ to yield $\mathcal{D}_{2/5}=\sqrt{5(\varphi^2 +1)}$, whereas for the non-Abelian $\nu=3/7$ state we may set $k=-3$, $M=3$ to yield $\mathcal{D}_{3/7}=\sqrt{7(\varphi^2 +1)}$. Note that in both cases this is larger than the minimum Abelian value for the total quantum dimension $\sqrt{s}$ by a factor of $\sqrt{\varphi^2+1}$. 

Since both the $\nu=2/5$ and $3/7$ non-Abelian theories have $|k|=3$, they describe a theory of Fibonacci anyons~\cite{Slingerland01}. As mentioned in the main text, the total quantum dimension must take the form $\mathcal{D}=\sqrt{\sum_a d_a^2}$, where $d_a$ is the quantum dimension of a quasiparticle of type $a$. For Abelian anyons $d_a=1$, whereas for non-Abelian anyons $d_a>1$~\cite{Kitaev06}. Furthermore, for any $\nu=r/s$ FQH state the ground-state degeneracy must be at least $s$, which confirms the minimum Abelian value for the total quantum dimension $\sqrt{s}$. In addition to Abelian anyons, a Fibonacci theory also hosts non-Abelian anyons of quantum dimension $\varphi$~\cite{Gils09}. Therefore, based on these two arguments alone, the total quantum dimension for a $|k|=3$ theory must take the form $\mathcal{D}_{|k|=3}=\sqrt{\alpha \varphi^2 +\beta}$, where $\alpha, \beta$ are positive integers satisfying $\alpha + \beta \geq s$.

\bibliographystyle{apsrev4-1}
\bibliography{tee}

\end{document}